\documentclass[12pt]{article}

\def\cB{{\cal B}}\def\cD{{\cal D}}
\def\cF{{\cal F}}\def\cG{{\cal G}}\def\cH{{\cal H}}
\def\cL{{\cal L}}

\def\cS{{\cal S}}\def\cT{{\cal T}}

\def\a{\alpha}

\def\l{\lambda}
\def\r{\rho}\def\s{\sigma}

\def\real{{\rm I\!R}}

\def\2#1{{}^{(2)}\!#1}\def\3#1{{}^{(3)}\!#1}
\def\4#1{{}^{(4)}\!#1}\def\+#1{{}^+\!#1}
\def\-#1{{}^-\!#1}
\def\*#1{{}^*\!#1}

\def\tr{\mathop{\rm Tr}}
\def\bra#1{\left\langle #1\right|}              
\def\ket#1{\left| #1\right\rangle}              

\newcommand{\braket}[2]{\langle #1 | #2 \rangle}

\def\diff{\rm\mathstrut d\!}

\newtheorem{teorema}{Theorem}
\newtheorem{prop}{Proposition}

\begin{document}
\title{{\bf Positive-Operator-Valued Time Observable in Quantum Mechanics}}
\author{R.Giannitrapani\thanks{e-mail:
riccardo@science.unitn.it} \\ {\normalsize\sl
Dipartimento di Fisica dell'Universit\`a di Trento}\\ {\normalsize\sl
I.N.F.N. gruppo collegato di Trento}}

\maketitle

\begin{abstract}
  We examine the longstanding problem of introducing a time observable
  in Quantum Mechanics; using the formalism of
  positive-operator-valued measures we show how to define such an
  observable in a natural way and we discuss some consequences.
\end{abstract}

Pacs: 03.65.Bz - Quantum theory;

UTF-390.

\section{Introduction}

Since the very beginning of Quantum Mechanics it has been clear that
it is not so easy to define time at a quantum level; in the ordinary
theory, in fact, it is not an observable, but an external parameter
or, that is the same, time is {\em classical}. In trying to change
this situation promoting time to be an observable, one has to face a
theorem by Pauli (Pauli 1958) that states, essentially, that such an
operator cannot be self-adjoint; since in the usual Quantum Mechanics
observables are postulated to be self-adjoint operators (see, for
example, Von Neumann 1955 and Prugove\v{c}ki 1971) this theorem
constitutes a problem.

One of the consequences of this is, for example, that one cannot
deduce the Heisenberg uncertainty relation for time and energy from a
kinematical point of view because time does not belong to the algebra
of observables. In spite of this the relation $\Delta T \cdot \Delta
H \geq 1$ is commonly accepted as true and it is derived in some way 
with dynamical considerations.

The situation is quite unsatisfactory both from a physical point of
view and from an epistemological point of view and although it has
been investigated in a good number of works (see, for example,
Aharonov et.al. 1961, Rosenbaum 1969, Olkhovsky et.al. 1974, Blanchard
et.al 1996, Grot et.al. 1996, Leon 1997), it is still an interesting
open problem.

The ``problem of time'' has some consequences also in the realm of
Quantum Gravity i.e. in the struggle to give a quantum description of
spacetime in order to solve some divergences problems in both General
Relativity (singularity theorems) and in Quantum Fields Theory
(renormalization problem). A quantum ``spacetime'' with zero spatial
dimensions and one time dimension (that is the quantization of time)
is the simplest model and we think it is preliminary to any other
attempt.

If one adopts the operational point of view (Bridgman 1927) then
defining the concept of time at a quantum level is equivalent to
specify a set of operations useful for the measurement of time; in
this context the problem of time is the problem of building ``quantum
clocks''. In this note we shall analyze a simple model for such a
quantum clock and try to draw some general conclusions on the problem.

\section{Mathematical preliminaries}

Our starting point is a generalized formulation of standard quantum
mechanics that extends the usual observable concept. A justification
of such formulation is given by Gleason's theorem (Bush et.al. 1991)
that guarantees that this structure is the most general one compatible
with the probabilistic interpretation of quantum mechanics (Copenhagen
interpretation); other justifications come from works by Ludwig and
Giles (Ludwig 1968, Giles 1970), but they are beyond the scopes of
this note.  In this section we summarize, in a very concise and
incomplete way, the mathematical tools that we shall use later; for a
good review of the subject, along with a very complete bibliography,
see Bush et.al. 1991, Giles 1970, Davies 1976.  \\

A given quantum system $\cS$ is described by an Hilbert space $\cH$;
we call $\cL(\cH)$ the algebra of bounded operators on $\cH$,
$\cL(\cH)^+$ the cone of positive ones and $\cT(\cH)$ the subalgebra
of the trace class operators.  The {\em states} of the system $\cS$
are the positive operators with trace one on $\cH$ that form a convex
set $\cT(\cH)^+_1$ in $\cT(\cH)$. \\

Given a measurable space $(\Omega,\cF)$, where $\Omega$ is a nonempty
set and $\cF$ a $\s$-algebra of subsets of $\Omega$, a {\em normalized
  positive operator valued measure} (a POV-measure) $\tau$ is a map

$$ \tau : \cF \rightarrow \cL(\cH)^+ $$
such that:

\begin{enumerate}
\item $\tau(X) \geq \tau(\O) = 0 \quad \forall X \in \cF$
\item $\tau(\cup X_i) = \sum \tau(X_i)$ where $\{X_i\}$ is a countable
  collection of disjoint elements of $\cF$ and the convergence is in
  the weak topology
\item $\tau(\Omega) = I.$
\end{enumerate}
If $\tau(X)^2 = \tau(X)$ than $\tau$ is a {\em projection valued
  measure} (PV-measure) and it can be demonstrated that this property is
equivalent to

$$\tau(X\cap Y) = \tau(X)\tau(Y).$$
If $\Omega$ is the real Borel space $(\real, \cB(\real))$ and 
$\tau$ is a PV-measure, than it is a spectral representation of a unique
self-adjoint operator $A$

\begin{equation} \label{adj1}
A = \int_\Omega \l \tau(\diff \l)
\end{equation}  

A {\em generalized observable} is a POV-measure on a particular
measurable space, while a PV-measure, via the relation (\ref{adj1}),
represents an ordinary observable of quantum mechanics. This
generalization of the concept of an observable is possible in view of
the probabilistic interpretation of quantum mechanics (for more
details see Bush 1991). Given an observable $\tau$ and a state $\r$
we have a probability measure $\tau_\r$ on $(\Omega,\cF)$

$$ \tau_\r : \cF \rightarrow [0,1] $$
$$ \tau_\r : X \mapsto \tr{[\r \tau(X)]}.$$ 
This can be interpreted as the probability that the measure of the
observable $\tau$ on the state $\r$ lies in the set $X$.

The mean value of the observable $\tau$ on the state $\r$ is then

$$ {\rm mean} (\tau,\r) = \int \l \tau_\r (\diff\l) $$
while the variance is given by

$$ {\rm var} (\tau,\r) = \int \l^2 \tau_\r (\diff\l) - ({\rm mean}
(\tau,\r))^2 . $$

Let $\cG$ be a locally compact group, $(\Omega,\cF)$ a
measurable $\cG$-space and $U$ a unitary representation 
of $\cG$ on an Hilbert
space $\cH$; if $\tau$ is a POV-measure on $(\Omega,\cF)$ with values in
$\cL(\cH)^+$, then we say that $\tau$ is {\em covariant} with respect to
$U$ if

$$ U_g \tau(X) U_g^* = \tau(X_g) $$ 
for every $X \in \cF$ and every $g \in \cG$. The pair $(\tau, U)$ is
called a {\em system of covariance} (Davies 1976); if $\tau$ is a
PV-measure then $(\tau, U)$ is a {\em system of imprimitivity}
(Mackey 1963, Varadarajan 1984).

The condition of covariance means that

$$ (U_g \tau(X) U^*_g)_\r = \tau_{\, U^*_g \r U_g}(X). $$

As it is stated in the introduction, due to an argument by Pauli
(Pauli 1958) it is not possible to have a self-adjoint operator for a
time observable in quantum mechanics; 

\begin{teorema}[Pauli] 
Given an observable (time) $T$ with the
following commutation relation with the hamiltonian

$$ [H,T] = -i $$
then $T$ cannot be a self-adjoint operator.
\end{teorema}
In the language of POV-measures the theorem means that a time
observable cannot form a system of imprimitivity with the time
translations, but it can still form a system of covariance with them.
In fact Pauli's Theorem is a consequence of the following general
proposition:

\begin{prop}
If $\tau$ is a POV-measure on $\real$ and it is covariant with respect
to the one parameter group of translation, then 

$$ \bra{\phi}\tau((a,b])\ket{\phi} > 0 \quad \forall \phi \in \cH $$
for every interval $(a,b]$;
this means that $\tau$ cannot be a PV-measures.
\end{prop}
{\bf\em Proof.}

For the demonstration of the proposition we can procede in the
following way: suppose that we have a POV-measure $\tau$ for the
observable time and that it forms a system of covariance with
$U=\exp{(-i\l H)}$, where $H$ is the generator of the
translations. Suppose that for a given pure state $\phi$ and a certain
interval of the real line $(a,b]$ we have

$$ \bra{\phi}\tau((a,b])\ket{\phi} = 0.$$
Then 

$$ \bra{\phi}\tau((a + \l, b - c + \l])\ket{\phi} = 0 \quad \forall \l
\in [0,c]. $$
and, for the covariance property,

$$\bra{\phi}e^{- i\l H} \tau((a,b-c]) e^{i\l H}\ket{\phi} = 0$$
and so

$$\bra{\phi} e^{- i\l H} \sqrt{\tau((a,b-c])} \sqrt{\tau((a,b-c])} 
e^{i\l H}\ket{\phi} = 0$$
for the positivity of $\tau$. At the end we have 

$$ F(\l) \equiv \sqrt{\tau((a,b-c])} e^{i\l H} \ket{\phi} = 0 \quad
\forall \l \in [0,c]. $$

But $F(\l)$ is an olomorphic vector valued function in the upper half
of the complex plane that is zero on the interval $[0,c]$; using the
{\em Riemann-Schwarz reflection principle} (Titchmarsh 1939) one can
prove that such a function, being zero on an interval, it is zero
everywhere. This means that $\bra{\phi}\tau((a + c -\l,b -
\l])\ket{\phi}$ is zero for all the values of $\l$
i.e. $\bra{\phi}\tau\ket{\phi}$ is zero on all the intervals of
$\real$ and this is impossible if $\tau$ has to be a normalized
POV-measure.

\begin{flushright}
Q.E.D.
\end{flushright} 

\section{A model for a Quantum Clock}
In this section we analyze a particular simple model for a quantum
clock (see Rosenbaum 1969, Toller 1996) using the mathematical
formalism of the preceding section.

Let us consider a one dimensional system represented by the Hilbert space

$$\cH = L^2 (\real).$$
We have, as usual, a coordinate $q$ observable along with its momentum
$p$ (in this case ordinary observables) such that

$$[q,p] =i.$$
Moreover this ``clock'' has an hamiltonian equal to

$$ H = \frac{p^2}{2}.$$
We can interpret $q$ as the time displayed by the clock and $p$ as the
rate of the clock itself.  In a classical model the real time would be

$$ T = \frac{q}{p},$$
but in the quantum case we have to take care of the ordering of the
operators. We have to perform an arbitrary choice and we follow
(Toller 1996) putting

$$ T = (2p)^{-1} q + q (2p)^{-1}. $$

This operator can be defined on the domain (in the
``$p$-representation) of infinitely differentiable functions over the
compact subsets of $\real - \{0\}$, that is dense in $\cH$ (it is also
possible to use as the domain the set of infinitely differentiable
functions over the compact subsets of $\real$ and then imposing the
condition of hermiticity that gives $ \displaystyle\lim_{p \rightarrow
0} \frac{\phi(p)}{\sqrt{p}} = 0 \quad \forall \phi \in \cD(T)$).

It is easy to see that $T$ is hermitean and the expected commutation
relation

$$ [H,T] = -i $$ 
is satisfied on $\cD(T)$. Now, for the Pauli theorem, we know that $T$
cannot be an ordinary observable, but we can still see if it can be
interpreted in the generalized framework of the preceding section. To
do so we have to find a POV-measure $\tau$ on $\real$ such that

$$ \bra{\phi} T \ket{\phi} = \int_\real \l \bra{\phi} \tau(\diff \l)
\ket{\phi} \quad \quad \forall \ket{\phi} \in \cD(T) \subset\cH.$$
Moreover $(\tau, U)$ has to be a covariance system with $U=\exp{(-i\l
H)}$ a representation of the time-translation group $\cG$.

In order to build $\tau$ let us start to search the eigenstates of
$T$; it is convenient to work in the momentum representation instead
of the usual coordinate representation (in such a way it is simpler to
define the operator $p^{-1}$). In such a representation we have

$$ T = i(2p)^{-1} \frac{\diff}{\diff p} + i\frac{\diff}{\diff
  p}(2p)^{-1}.$$

The eigenvector problem reads as
$$ T \ket{t} = t \ket{t}$$
and defining the wavefunction $\psi_t(p)$ as
$$ \psi_t(p) =  \braket{p}{t} $$
we have
$$ T\, \psi_t(p) = t \psi_t(p).$$ 
This equation admit as solutions a double family of eigenfunctions:

$$\braket{p}{t,\a} = \psi_{t\a}(p) = \frac{1}{\sqrt{2\pi}}
  \theta(\a p)\, \sqrt{|p|}\,
  \exp{\left(-\frac{itp^2}{2}\right)}$$ 
with $\a = \pm 1$.
They do not lie in $\cH$ and so they have to be regarded as weak
eigenfunctions:

$$ \bra{t,\a}(T-t)\ket{\phi} = 0 \quad \forall \phi \in \cD(T).$$

We can also see easily that the eigenvectors of $T$ are not orthogonal

$$ \braket{t,\a}{t',\a'} = 0  \quad {\rm with}\, \a \neq \a' $$
$$ \braket{t,\a}{t',\a} = \frac{1}{2}\delta
(t-t') + \frac{i}{2} P\frac{1}{\pi (t-t')}. $$

Anyway the following relation still holds (in the weak sense):

$$ \sum_\a \int_{-\infty}^{+\infty} dt \ket{t,\a}\bra{t,\a} = {\bf
1}.$$\\

At this point we can state the following propositions:

\begin{prop}
  $\tau (dt) = \sum_\a \ket{t,\a} \bra{t,\a} \diff t$ gives a POV-measure

  $$\tau(X) = \int_X \tau(dt) = \sum_\a \int_X
  \ket{t,\a}\bra{t,\a}\diff t $$ with $X$ a Borel set of the real
  line.
\end{prop}

\begin{prop}
  The system $(\tau, U)$, where $U=\exp{(-i\l H)}$ is a representation
  of the one parameter group $\cG$ of time translations, is a
  covariance system.
\end{prop}
{\bf\em Proof.}

Let us start from the first one; obviously $\tau(X)$ is a positive
operator, moreover it is bounded

$$ \tau(X) \leq \tau(\real) = \sum_\a\int_{-\infty}^{+\infty} \diff t
\ket{t,\a}\bra{t,\a} = {\bf 1}$$
so that $\tau(X) \in \cL^+(\cH)$. The $\sigma$-additivity follows from
the additivity of integrals and $\tau$ is normalized to $\bf 1$.

For the second proposition one can see that

$$ e^{i\l H} \ket{t,\a} = \ket{t - \l,\a} $$
and so

\begin{eqnarray*}
\bra{\phi} e^{i\l H} \tau(X) e^{-i\l H} \ket{\phi} & = & \int_X
\bra{\phi} e^{i\l H} \tau(\diff t) e^{-i\l H} \ket{\phi} = \\ & = &
\sum_\a \int_X \diff t \bra{\phi} e^{i\l H} \ket{t,\a} \bra{t,\a}
e^{-i\l H} \ket{\phi} = \\ & = & \sum_\a \int_X \diff t
\braket{\phi}{t-\l,\a} \braket{t-\l,\a}{\phi} = \\ & = & \bra{\phi}
\tau(X-\l) \ket{\phi}
\end{eqnarray*}
that is the relation of covariance of the POV-measure $\tau$.

\begin{flushright}
Q.E.D.
\end{flushright}

In conclusions we can say that $\tau$ is a generalized observable for
the time of our quantum clock; it can be checked that $\tau$ is not a
PV-measure (essentially this is a consequence of the non orthogonality
of the eigenvectors of $T$) and so there is no contradiction with the
Pauli Theorem.

We have studied a particular POV-measure for a time observable
obtained by choosing a very particular time operator; the next step is
to study POV-measures for time regardless of any operator. The
interesting object is the space of POV-measures that form a system of
covariance with a representation of time translations; the task is to
find out in such a space the ``best'' measures to be used for quantum
clocks. This will be the argument of a future note.

\section{Uncertainty Relations}

We now can examine the uncertainty relations for time and energy
from a kinematical point of view, as stated in the introduction.

If we define, for an hermitean operator $A$, the quantity

$$ (\sigma_A)^2 = \bra{\phi} A^2 \ket{\phi} -
(\bra{\phi}A\ket{\phi})^2 \quad {\rm with}\,\, \braket{\phi}{\phi} = 1
$$
then one can prove (Von Neumann 1955) that for the operators $T$ and
$H$ of the preceding section (they are hermitean) the following
relation is true on a certain domain of $\cH$

$$ \sigma_T \sigma_H \geq \frac{1}{2}. $$ 
This relation is commonly accepted as the equivalent for time and
energy of the famous Heisenberg relation for position and momentum;
the fact is that the quantity $\sigma_T$ is not, in general, the
variance of the observable time $\Delta T$ because $T$ is a
generalized observable and it is not a self-adjoint operator. But in
our simple model the two quantities coincide; in fact we can write

$$ T \ket{\phi} = \sum_\a \int_{-\infty}^{+\infty} \diff t
        \ket{t,\a}\bra{t,\a} T \ket{\phi} \quad \forall \ket{\phi} \in
        \cD(T) $$

for the property of $\tau$ exposed in the preceding section; since
$\ket{t,\a}$ is a weak eigenvector of $T$ we have 

$$ T \ket{\phi} = \sum_\a \int_{-\infty}^{+\infty} \diff t\, t\, 
\ket{t,\a}\braket{t,\a}{\phi}. $$
From this relation one sees that the mean of $T$, as defined
in the second section, is the usual one

$$ \bra{\phi} T \ket{\phi} = \sum _\a \int_{-\infty}^{+\infty} \diff
t\, t \braket{\phi}{t,\a}\braket{t,\a}{\phi} =
\int_{-\infty}^{+\infty} t \, \bra{\phi}\tau(\diff t) \ket{\phi} =
{\rm mean}(\tau,\phi). $$

Using the relation

$$ \bra{\phi} T^2 \ket{\phi} = \sum_{\a} \sum_{\a'}
\int_{-\infty}^{+\infty} \diff t \int_{-\infty}^{+\infty} \diff t'\, t
t'\, \braket{\phi}{t',\a'}\braket{t',\a'}{t,\a}\braket{t,\a}{\phi} $$
we obtain

$$ \bra{\phi} T^2 \ket{\phi} = \frac{1}{2}\int_{-\infty}^{+\infty} t^2
\bra{\phi}\tau(\diff t) \ket{\phi} + \Lambda_\phi$$ 
where

$$ \Lambda_\phi = \frac{i}{2\pi} \sum_\a \int_{-\infty}^{+\infty}
\diff t \int_{-\infty}^{+\infty} \diff t' \,
\overline{t'\phi(t',\a)}\, t\phi(t,\a) \, P\frac{1}{(t' - t)}$$ with
$\phi(t,\a) = \braket{t,\a}{\phi}.$ 
One can check that

$$ \Lambda_\phi = \frac{1}{2}\int_{-\infty}^{+\infty} t^2
\bra{\phi}\tau(\diff t) \ket{\phi}$$ 
and than 

$$ \bra{\phi} T^2 \ket{\phi} = \int_{-\infty}^{+\infty} t^2
\bra{\phi}\tau(\diff t) \ket{\phi}.$$ 
In the end we have for the generalized observable $\tau$

$$ \sigma_T = {\rm var}(\tau,\phi) $$ 
and so the uncertainty relation for time and energy variances is
obtainable in a
rigorous way within the POV-measures formalism. 

\section{Conclusions}

In this note we have shown how it is possible to give a well defined
meaning to the concept of time observable at a quantum level using the
POV-measures formalism; in particular we have studied a simple quantum
clock model giving a precise mathematical derivation of the Heisenberg
uncertainty relation for time and energy.  Since clocks are
fundamental in the operational definition of spacetime, in our mind
this is a preliminary step toward an analysis of spacetime concepts at
a quantum level, analysis that we hope to present in future works.

\subsection*{Aknowledgments}

I would like to express my sincere gratitude to M.Toller for his
encouragement and for suggestions without which this work would not
have been completed. I wish also to thank V.Moretti for the help in
solving some technical problems and for useful discussions.  I would
like to thank H.Atmanspacher and W.M. de Muynck for having pointed out
to me, after the acceptance of this paper, a work of Busch {\em et
  al.} (1994) and of Holevo (1982) concerning the same topic.

\newpage

\noindent {\bf \Large References}
\vspace{0.5cm}
\begin{description}

\item Aharonov, Y., and Bohm, D. (1961). {\rm Time in the Quantum
Theory and the Uncertainty Relation for Time and Energy}, {\em
Physical Review}, {\bf 122} 1649.

\item Blanchard, Ph., and Jadczyk, A. (1996). {\rm Time of Events in
Quantum Theory}, {\em preprint quant-ph/9602010}.

\item Bridgman, P.W. (1927). {\em The Logic of Modern Physics},
  The Macmillan Company, New York.

\item Busch, P., Lahti, P.J., and Mittelstaedt, P. (1991). {\em The Quantum
    Theory of Measurement}. Lectures Notes in Physics m2, Springer Verlag,
  Berlin.

\item Busch, P., Grabowski, M., and Lathi, P.J. (1994). Time
  Observables in Quantum Theory, {\em Physics Letters A}, {\bf 191},
  357. 

\item Davies, E.B. (1976). {\em Quantum Theory of Open Systems},
  Academic Press, London.

\item Giles, R. (1970). {\rm Foundations for Quantum Mechanics},
  {\em Journal of Mathematical Physics}, {\bf 11} 2139.

\item Grot, N., Rovelli, C., and Tate, R.S. (1996). {\rm Time-of-arrival in
    quantum mechanics}, {\em preprint quant-ph/9603021}.

\item Holevo, A.S. (1982). {\rm Probabilistic and statistical aspects
    of quantum theory}, North-Holland Publ, Amsterdam.

\item Leon, J. (1997). {\rm Time-of-arrival formalism for the
    relativistic particle.} {\em Journal of Physiscs} {\bf A30}, 4791.

\item Ludwig, G. (1968). {\rm Attempt of an Axiomatic Foundation of
Quantum Mechanics and More General Theories III.}  {\em
Communications in Mathematical Physics},  {\bf 9} 1.

\item Mackey, G.W. (1963). {\rm Infinite Dimensional group
        representations}, {\em Bulletin of the American Mathematical
        Society}  {\bf 69}, 628.

\item Olkhovsky, V.S., Recami, E., and Gerasimchuk, A.J. (1974). {\rm
    Time Operator in Quantum Mechanics. I: Nonrelativistic Case},
    {\em Nuovo Cimento}, {\bf 22}  263.

\item Pauli, W. (1958). {\em Die allgemeinen Prinzipien der
    Wellenmechanik},  Hanbuck der Physik, edited by S.Fl\"ugge,
    vol.  V/1, p.60, Springer Verlag, Berlin.
           
\item Prugove\v{c}ki, E. (1971). {\em Quantum Mechanics in Hilbert
    Space}, Academic Press, New York and London.

\item Rosenbaum, D.M. (1969). {\rm Super Hilbert Space and the Quantum
    Mechanical Time Operators}, {\em Journal of Mathematical
    Physics}, {\bf 10} 1127.

\item Titchmarsh, E.C. (1939). {\em The Theory of Functions},
        Oxford University Press, Oxford.

\item Toller, M. (1996). {\rm Quantum References and Quantum
    Transformations}, {\em preprint gr-qc/9605052}.

\item Varadarajan, V.S. (1984). {\em Geometry of Quantum Theory} (Second
    Edition), Springer Verlag, Berlin.

\item Von Neumann, J. (1955). {\em Mathematical Foundations of Quantum
    Mechanics}, Princeton University Press, Princeton.

\end{description}

\end{document}